# ІНФОРМАЦІЙНІ ТЕХНОЛОГІЇ, СИСТЕМНИЙ АНАЛІЗ ТА КЕРУВАННЯ

S.M. Alkhimova*

Igor Sikorsky Kyiv Polytechnic Institute, Kyiv, Ukraine

## AUTOMATED DETECTION OF REGIONS OF INTEREST FOR BRAIN PERFUSION MR IMAGES

**Background.** Images with abnormal brain anatomy produce problems for automatic segmentation techniques, and as a result poor ROI detection affects both quantitative measurements and visual assessment of perfusion data.
**Objective.** This paper presents a new approach for fully automated and relatively accurate ROI detection from dynamic susceptibility contrast perfusion magnetic resonance and can therefore be applied excellently in the perfusion analysis.
**Methods.** In the proposed approach the segmentation output is a binary mask of perfusion ROI that has zero values for air pixels, pixels that represent non-brain tissues, and cerebrospinal fluid pixels. The process of binary mask producing starts with extracting low intensity pixels by thresholding, which subsequently correspond to zero values of the mask. Optimal low-threshold value is solved by obtaining intensity pixels information from the approximate anatomical brain location. Holes filling algorithm and binary region growing algorithm are used to remove falsely detected regions and produce region of only brain tissues. Further, CSF pixels extraction is provided by thresholding of high intensity pixels from region of only brain tissues. Each time-point image of the perfusion sequence is used for adjustment of CSF pixels location.
**Results.** The segmentation results were compared with the manual segmentation performed by experienced radiologists, considered as the reference standard for evaluation of proposed approach. On average of 120 images the segmentation results have a good agreement with the reference standard with a Dice Index of $0.9576 \pm 0.013$ (sensitivity and specificity are $0.9931 \pm 0.0053$ and $0.9730 \pm 0.0111$ respectively). All detected perfusion ROIs were deemed by two experienced radiologists as satisfactory enough for clinical use.
**Conclusions.** The results show that proposed approach is suitable to be used for perfusion ROI detection from DSC head scans. Segmentation tool based on the proposed approach can be implemented as a part of any automatic brain image processing system for clinical use.

**Introduction**

Dynamic susceptibility contrast (DSC) perfusion magnetic resonance (MR) imaging is used increasingly in clinical practice to evaluate cerebral blood perfusion in patients with such diseases as carotid stenosis, stroke or brain tumors [1]. In this technique, the passage of contrast agent particles through the entire brain causes signal drop (shortening of T2 and T2*) within tissues. This effect is monitored by a dynamic series of MR images. The recorded images are processed on a pixel-by-pixel basis to provide quantification of different hemodynamic parameters and to generate color-coded perfusion maps for them. Therefore analysis of DSC perfusion data can be based on quantitative perfusion measurements or on visual interpretation of perfusion maps. However, poor placement of perfusion region of interest (ROI) affects both quantitative measurements and visual assessment of perfusion data [2, 3]. This can be explained by the fact of presence of numerous artifacts that are obtained from image noise or from such non-brain tissues as skull or from regions filled with cerebrospinal fluid (CSF) [4]. Most of the software programs for perfusion data analysis provide the possibility to detect perfusion ROI in manual mode or could be more or less automated. Automatic approaches for segmentation are preferable in clinical practice for few reasons. First of all, manual segmentation is a very time-consuming process that can be critical for such patients that required acute stroke treatment. Furthermore, operator knowledge on brain structures and lesions is essential for correct results. Although the process of manual segmentation is a subjective process and is not reproducible. In spite of the above benefits, automatic segmentation approaches are generally good enough only for slices with normal anatomy. Image regions with lesions (tumor, stroke, necrosis, etc) can be lost because of automatic segmentation fails, and it is unacceptable for perfusion analysis. It can be explained by grounds of widely used algorithms for such automatic segmentation. Whereas there are a wide variety of proposed approaches, most of them can be grouped in two classes. One of them is based on image intensity values (thresholding, clustering algorithms), and the other is based on pattern recognition (neural network classifier, atlas-based algorithms) [5, 6]. The main drawback of the intensity based approaches is the problem of overlapping pixel intensity in lesion tissues and background. Lack of age-sex-race-specific pre-segmented template data as well as not enough training samples for different size, density, and volume of the lesion can be a drawback for pattern based approaches. Also, it should be mentioned that most automated segmentation approached are focused on T1 images and may not be suitable for T2 images. Proposed strategy with parameterization of the T2 image intensity onto a standardized T1 intensity [7] scale uses mentioned before approaches for segmentation process, thus it is subject to the same drawbacks. There are some other approaches for head MR-scans segmentation that use benefits from specific image acquisition technique, but they are out of scope of this research paper.

* corresponding author: asnarta@gmail.com

**Problem statement**

The purpose of this study is to propose a fully automated, simple in implementation, and fast in execution time segmentation approach for perfusion ROI detection from DSC head scans. Threshold-based pixel differentiation forms the basis of this approach. Approximate anatomical brain location is uses to obtain information for solving optimal threshold value (its detection is based on both intensity features extraction and edge detection from derivatives of image projection profiles).

The rest of the paper is organized as follows. Section 2 discusses the proposed materials and methods. Section 3 presents and discusses the segmentation results. Finally, section 4 concludes the paper.

**Material and methods**

All brain perfusion MR images were acquired on a 3.0 T clinical MR scanner (Achieva, Philips Healthcare, Best, the Netherlands) from 12 patients with cerebrovascular disease. MR imaging was performed by using a T2*-weighted echo-planar sequence. Scan parameters were: repetition time, 1500 ms; echo time, 30 ms; flip angle, 90°; field of view, $23 \times 23$ cm; image matrix, $256 \times 256$; slice thickness, 5 mm; gap, 1 mm. 17 slices were scanned with 40 dynamic images for each slice. Contrast medium (Gadovist, Bayer Schering Pharma AG, Berlin, Germany) with a dosage of 0.1 mmol/kg body weight was injected at a rate of 5 mL/sec, followed by a 10-mL bolus of normal saline also at 5 mL/sec. All images were collected in 12-bit DICOM (Digital Imaging and Communication in Medicine) format.

Image postprocessing software program was in-house developed. It is written in C++ and uses an open-source The Grassroots DiCoM library (GDCM, http://gdcm.sourceforge.net/) for loading medical images. It uses multi-threaded implementation to speed-up the data processing.

In the proposed approach the segmentation output is a binary mask of perfusion ROI that has zero values for air pixels, pixels that represent non-brain tissues, and CSF pixels. Initially a binary mask has unity value for all image pixels. There are two major steps to produce binary mask of perfusion ROI from the initial all-unity mask: (1) low intensity pixels extraction; (2) high intensity pixels extraction.

The first step of the segmentation process consists in removing low intensity pixels from the entire image, i.e. air pixels and pixels of non-brain tissues (skull and extracranial soft tissues). This step reduces unity values of the initial binary mask to only brain tissues pixels.

The simplest and most commonly used method for such procedure is thresholding. It differentiates foreground (in our case brain tissues pixels) from the background (air and non-brain tissues pixels) through finding threshold value and marking image pixels with higher-than-threshold intensities as one class and the rest as another class. Finally, based on marking results, a binary image is created by assigning a value of zero to background pixels and a value of unity to foreground pixels.

Finding an optimal threshold value is a key point to obtain a good segmentation quality. In case of threshold value is derived from the intensity statistics of the entire image, commonly applied methods such as Otsu, entropy-based or thresholding based on standard deviation yield an incorrect threshold value and fail to identify required low intensity pixels. In order to keep more hypointense pixels of brain tissues and lesions and at the same time to avoid involving of noised air pixels and pixels of non-brain tissues to final foreground results, the proposed approach determines low-threshold value as a derived from approximate anatomical brain location. Approximate anatomical brain location indicating the probable area of the brain and it is obtained from the analysis of image projection profiles as follows.

Projection profile technique is widely used for indicating approximate anatomical positions of some organs or tissues [8, 9]. Different from commonly used averaged projection the proposed approach uses projection profile that is 1D function of the standard deviation values obtained by projecting the image pixels onto horizontal or vertical axis.

Let $I[x,j]$ denote the pixel intensity value at $(x, y)$ coordinates. Then, the horizontal standard deviation projection, denoted by $P_H[x]$, is defined as

$$P_H[x] = \sigma[x] = \sqrt{\frac{\left(\sum_{j=0}^{M-1} I^2[x,j] - \frac{1}{N}\left(\sum_{j=0}^{M-1} I[x,j]\right)^2\right)}{(N-1)}}$$

where $N$ is the number of image columns and $M$ is the number of image rows. The vertical standard deviation projection $P_V[x]$ is defined the same way by projecting the image pixels onto vertical axis and using them as a sample to calculate the standard deviation value.

The standard deviation value gives a measure of pixels intensity scattering across each projection. Projections with a lower standard deviation will have more uniform pixel intensities throughout its pixels. Therefore projection throughout the region with air pixels shows the lowest standard deviation values. Projection throughout the region with a lot of uniform intensity from air pixels and with outlier of different intensity from skull pixels will still show a low standard deviation value. As the projection will capture brain tissues pixels standard deviation value will become significantly greater. The global extrema of the first derivative of the projection curves are taken to be the edges of the approximate anatomical brain location. This is the basic idea of the entire image cropping to extract area of the approximate anatomical brain location. An example of the projection profiles of an image is shown in Fig. 1. In order for optimal results to be achieved, extracting of mentioned area is performed on the 4[th] time-point image for each space position (image acquisition protocol can vary, so processed image should be selected after discarding the first few time-points images at which signal intensity is not reached a steady state).

Intensity pixels information from the approximate anatomical brain location is used to determine low-threshold value. The low-threshold value $T_L$ is calculated as

$$T_L = \mu_L - \sigma_L$$

where $\mu_L$ and $\sigma_L$ are mean and standard deviation of intensity values for all the pixels included in the region of approximate anatomical brain location.

Sometimes a binary image that is obtained by applying thresholding with found low-threshold value $T_L$ has falsely detected regions: small holes located within the foreground, i.e., hypointense areas inside the region of brain tissues (falsely detected background regions), and/or small isolated areas located in places of non-brain tissues, i.e. skull, skin, muscles, fat, etc. (falsely detected foreground regions). This happens due to the overlapping range of intensity for low intensity objects (intracranial tissues and lesions with low signal intensity on analysed images) and background.

In order to improve the binary mask of only brain tissues by removing falsely detected background regions a hole filling algorithm is applied [10]. The hole filling process is based on marking of each isolated background region as true background or as a hole in accordance with the following condition: the region is a background if its pixels end with image limit, otherwise it is a hole. Then all found holes are removed by changing their region with foreground.

After holes filling, a binary region growing algorithm is used to find the largest connected region, which is assumed to be a region that identifies only brain tissues [11].

The binary mask of only brain tissues with intermediate processing steps to obtain it through low intensity pixels extraction are shown in Fig. 2.

The second step of the segmentation process consists in removing high intensity pixels from previously detected region of brain tissues, i.e. CSF pixels. This step reduces unity values of existing binary mask more to produce final mask of perfusion ROI. The suppression of CSF inside the ventricles allows better assessment of tissues with delayed perfusion on such maps as time to peak or mean transit time that notably contaminated by hyperintense CSF pixels in case of their presence.

Similar to the first step of low intensity pixels extraction, thresholding is also can be used as a simple and fast method to identify CSF pixels that have high intensity difference from the surrounding pixels. However, in common cases thresholding leads to the identification of not only CSF pixels. Brain lesions nature can result in overlapping of pixels intensities range in lesion tissues and regions filled with CSF on DSC perfusion MR images. It is the main reason of failing to have correct high intensity pixels extraction. Binary images with good CSF pixels coverage can be derived by applying thresholding to the ratio images [12], but there are a number of false positive segmented regions (mostly belonging to the lesions tissues) on them as well. In order to reduce false positive regions detection and at the same time to identify as more as possible CSF pixels for further removing, the proposed approach determines high-threshold value as a derived from region of only brain tissues and uses each time-point image of the perfusion sequence for adjustment of CSF pixels location.

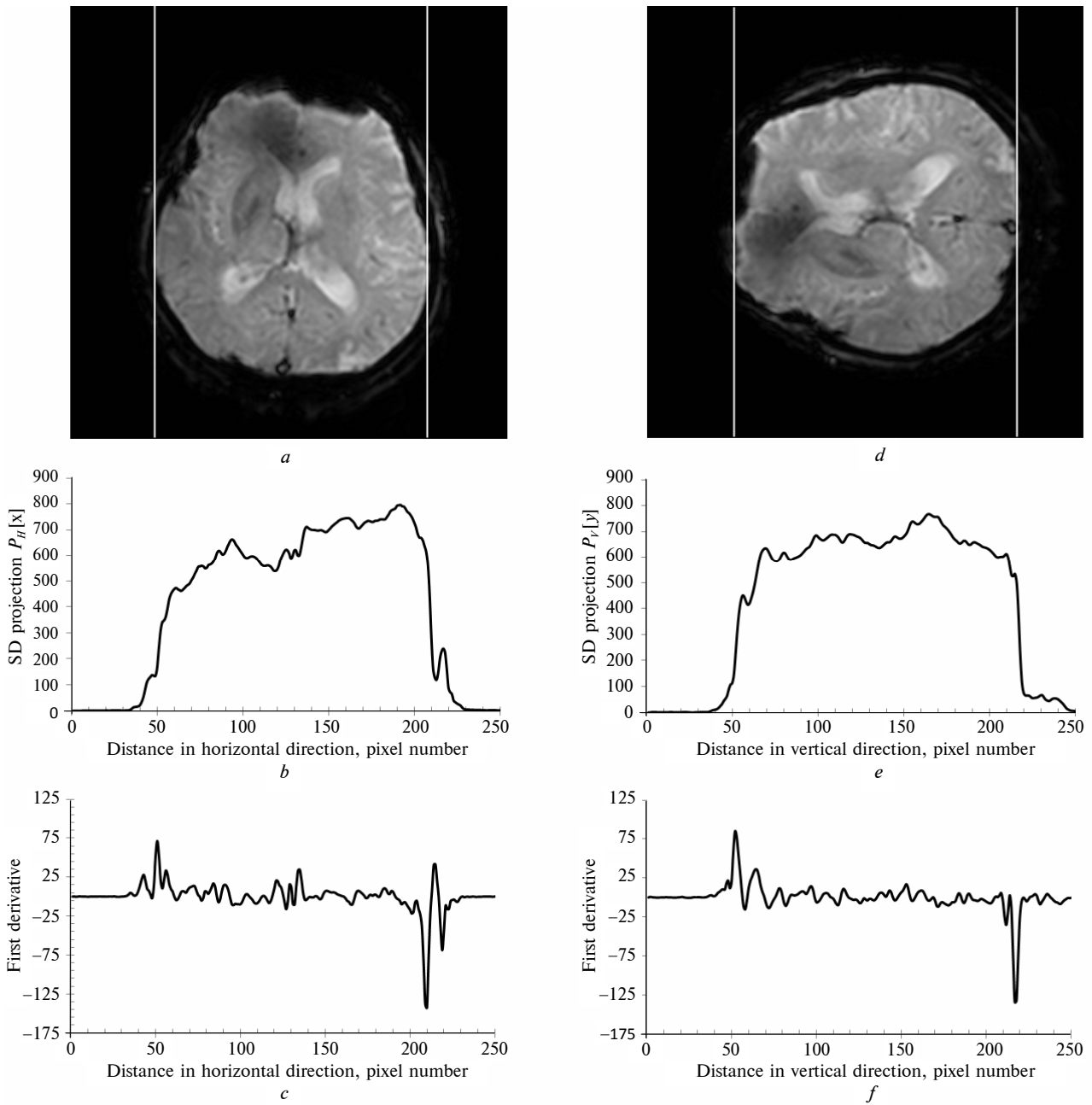

Fig. 1. (*a*) Sample of image with horizontal edges of the approximate anatomical brain location. (*b*) Horizontal standard deviation projection profile of *a*. (*c*) First derivative of the projection curve in *b*. (*d*) The same image as *a* with vertical edges of the approximate anatomical brain location. (*e*) Vertical standard deviation projection profile of *d*. (*f*) First derivative of the projection curve in *e*

High-threshold value $T_H$ is determined based on the intensity pixels information from the obtained on the first step region of only brain tissues. It is calculated as

$$T_H = \mu_H + \sigma_H$$

where $\mu_H$ and $\sigma_H$ are mean and standard deviation of intensity values for all the pixels included in the region of only brain tissues.

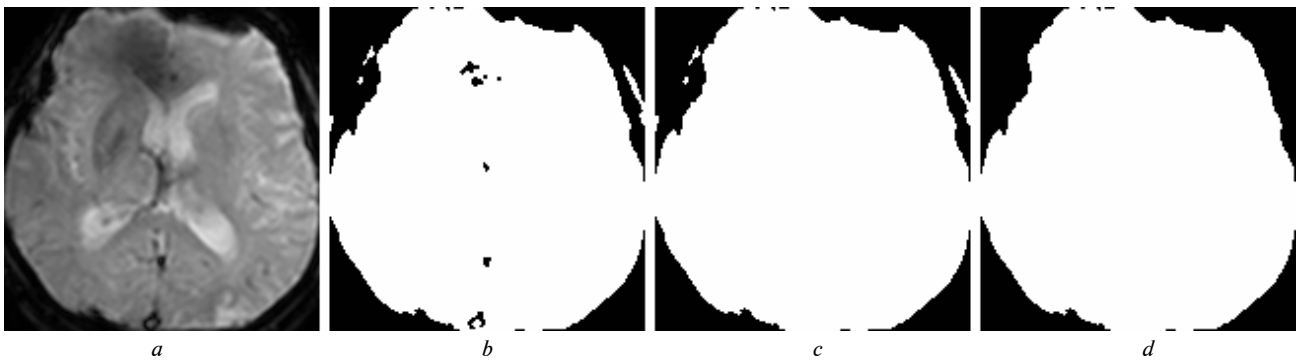

Fig. 2. (*a*) Approximate anatomical brain location area for sample image from Fig. 1. (*b*) Binary image produced using thresholding on *a* with found low-threshold value $T_L$. (*c*) Results of removing interior holes in the binary image from *b*. (*d*) Results of finding binary mask of only brain tissues as largest connected component on binary image from *c*

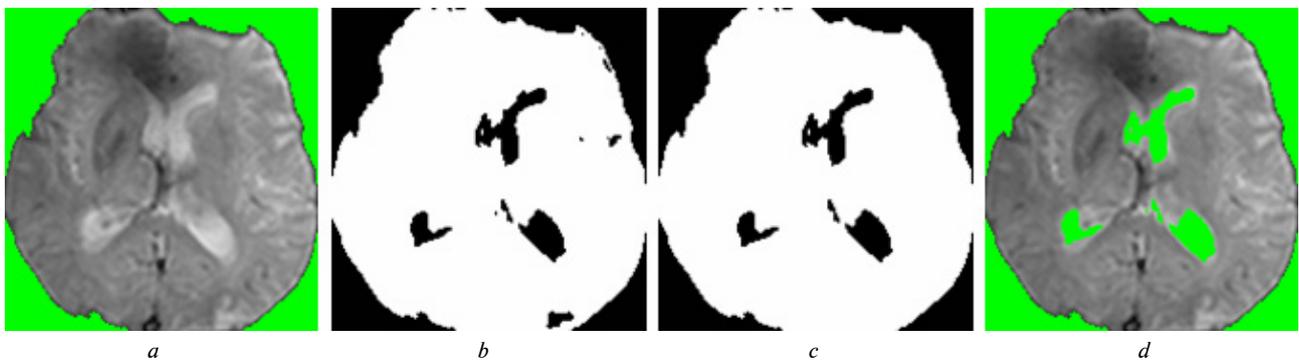

Fig. 3. (*a*) Brain region without skull and extracranial soft tissues on the 4th time-point image extracted with binary mask of only brain tissues from Fig. 2. (*b*) Binary image produced using thresholding of *a* with found high-threshold value $T_H$. (*c*) Results of finding binary mask of perfusion ROI as intersecting of binary images obtained from all time-points images with the same space location as *a*. (*d*) Perfusion ROI extracted with binary mask from *c* for quantitative measurements and visual assessment of brain perfusion data

Found high-threshold value $T_H$ is applied to each time-point image of the registered perfusion sequence to produce series of binary images. Each binary image is created from the existing binary mask of only brain tissues by assigning to them zero values in places where image pixels have greater intensities than $T_H$. It is able to make CSF pixels identification by preliminary reducing processing area that is close to the ventricle system. It can be effective to improve processing speed.

Adjustment of the CSF pixels location is performed by intersection of all obtained through applying $T_H$-thresholding binary images that have the same space location. Produced through these actions mask is a final binary mask of perfusion ROI. This mask is applied on given image to extract the perfusion ROI from image.

The binary mask of perfusion ROI with intermediate processing steps to obtain it through high intensity pixels extraction, and extracted with this mask perfusion ROI are shown in Fig. 3.

### Results and discussion

The proposed segmentation approach was applied to detect perfusion ROI on DSC head scans from 12 clinical cases. The segmentation results obtained from the proposed approach were compared with a reference standard, which is the manually marked ROI of the brain perfusion data by an experienced radiologist and confirmed by a second radiologist. 10 sample images were selected from each of the clinical cases as the test set for the validation experiments.

In order to evaluate the accuracy of the proposed approach area-based metric was considered to estimate the similarity between the segmented perfusion ROIs through the proposed approach $(ROI_P)$ and reference standard $(ROI_R)$. Dice In-dex $DI = 2|ROI_P \cap ROI_R|/(|ROI_P|+|ROI_R|)$ was used to estimate spatial overlap between two ROIs. It values range between 0 (no overlap) and 1 (perfect agreement).

Sensitivity and specificity were used to estimate the number of properly detected pixels. The sensitivity shows the ability to correctly detect pixels of perfusion ROI. It is defined as True Positive Fraction $TPF = TP/(TP+FN)$, which calculation is based on true positive ($TP$) and false negative ($FN$) regions. The higher sensitivity shows the lower missed true pixels of perfusion ROI. The specificity shows the ability to correctly detect pixels of background (in our case air pixels, pixels that represent non-brain tissues, and CSF pixels). It is defined as True Negative Fraction $TNF = TN/(TN+FP)$, which calculation is based on true negative ($TN$) and false positive ($EP$) regions. The higher specificity shows the lower missed true pixels of background.

Evaluation results of the proposed segmentation approach are shown in the Table.

The results show that the proposed approach can reliably detect the perfusion ROI from DSC head scans. On average of 120 images the segmentation results have a good agreement with the reference standard with a Dice Index of $0.9576 \pm 0.0130$. Furthermore, the sensitivity and specificity measures of all the clinical cases are high ($0.9931 \pm 0.0053$ and $0.9730 \pm 0.0111$ respectively). This means that detected regions are very satisfactory for quantitative measurements and visual assessment of brain perfusion data and do not need to be manually edited much by radiologist. All detected perfusion ROIs were deemed by two experienced radiologists as satisfactory enough for clinical use.

Comparison of the proposed approach results with user-defined threshold method was done on the same database. User-defined thresholding is the state-of-the-art method for perfusion ROI detection that is used in brain image processing system

*Table.* Evaluation results (average value and standard deviation) of perfusion ROI detection from DSC head scans

| Clinical case | Metrics | | |
|---|---|---|---|
| | Similarity (DI) | Sensitivity (TPF) | Specificity (TNF) |
| 1 | $0.9835 \pm 0.0118$ | $0.9969 \pm 0.0020$ | $0.9863 \pm 0.0063$ |
| 2 | $0.9704 \pm 0.0293$ | $0.9973 \pm 0.0003$ | $0.9760 \pm 0.0164$ |
| 3 | $0.9361 \pm 0.0124$ | $0.9985 \pm 0.0012$ | $0.9557 \pm 0.0204$ |
| 4 | $0.9382 \pm 0.0141$ | $0.9733 \pm 0.0173$ | $0.9655 \pm 0.0099$ |
| 5 | $0.9714 \pm 0.0156$ | $0.9970 \pm 0.0015$ | $0.9800 \pm 0.0105$ |
| 6 | $0.9809 \pm 0.0103$ | $0.9989 \pm 0.0010$ | $0.9856 \pm 0.0104$ |
| 7 | $0.9727 \pm 0.0097$ | $0.9972 \pm 0.0018$ | $0.9812 \pm 0.0182$ |
| 8 | $0.9801 \pm 0.0105$ | $0.9992 \pm 0.0007$ | $0.9836 \pm 0.0110$ |
| 9 | $0.9817 \pm 0.0077$ | $0.9969 \pm 0.0029$ | $0.9857 \pm 0.0054$ |
| 10 | $0.9007 \pm 0.0107$ | $0.9790 \pm 0.0181$ | $0.9511 \pm 0.0111$ |
| 11 | $0.9026 \pm 0.0128$ | $0.9834 \pm 0.0160$ | $0.9458 \pm 0.0078$ |
| 12 | $0.9725 \pm 0.0112$ | $0.9991 \pm 0.0007$ | $0.9793 \pm 0.0062$ |

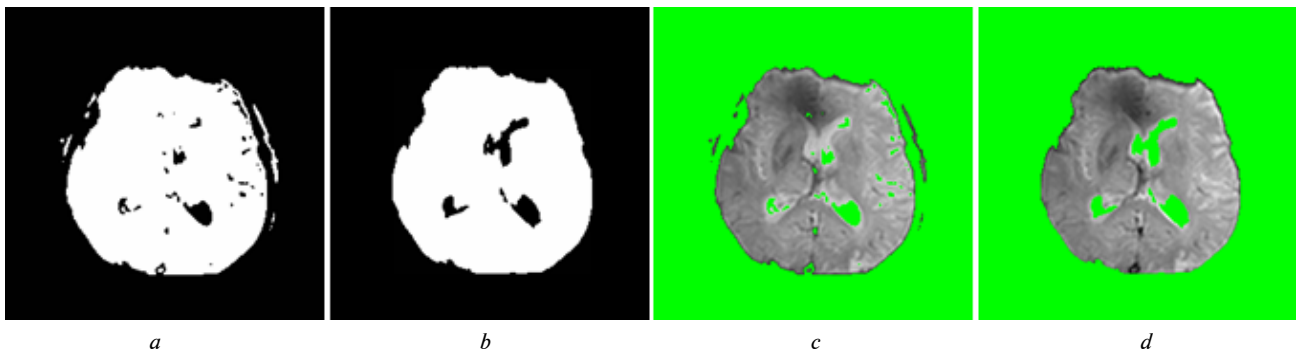

Fig. 4. (*a*) Binary mask of perfusion ROI produced with using of user-defined threshold method for sample image from Fig. 1. (*b*) Binary mask of perfusion ROI produced with using of proposed in current study approach for sample image from Fig. 1. (*c*) Perfusion ROI extracted with binary mask from *a*. (*d*) Perfusion ROI extracted with binary mask from *b*

for clinical use. More details on automation, optimal threshold selection, and obtained results for user-defined threshold method can be found in previous study [2]. Proposed in current study approach showed superior performance compared with user-defined threshold method that reached only to 0.7829 ± 0.0866 for evaluating with Dice Index metric. An example of segmentation result for sample image is shown in Fig. 4.

As shown in the results, proposed approach of perfusion ROI detection from DSC head scans presents significant accuracy. In addition, one should note that proposed approach of perfusion ROI detection should be tested on more data sets obtained with different scan parameters in order to be better validated.

Average processing time of the proposed approach is about 0.2s per image (Intel Core i7-3770 3.4GHz, single threaded), which is quite fast.

### Conclusions

Segmentation approach for perfusion ROI detection from DSC head scans was proposed in this paper. It does not require any external parameter to initialize image processing and thus to be considered as a fully automated. The proposed approach uses threshold-based pixel differentiation and solves optimal threshold value through processing pixels from detected approximate anatomical brain location. The evaluation results show that the strength of proposed approach is in producing well enough perfusion ROI detection from DSC head scans for slices with abnormal brain anatomy and, at the same time, it does not require a large computation time.

In summary, the approach proposed in this study provides accurate enough perfusion ROI automated segmentation on DSC head scans as suggested by evaluation analysis on images from 12 clinical cases. Proposed approach is fully automated, thus it can play an important role in cases that require processing of a large amount of patient images. Segmentation tool based on the proposed approach can be implemented as a part of any automatic brain image processing system for clinical use.

С.М. Алхімова

АВТОМАТИЧНЕ ВИЗНАЧЕННЯ ЗОН УВАГИ НА МАГНІТНО-РЕЗОНАНСНИХ ПЕРФУЗІЙНИХ ЗОБРАЖЕННЯХ МОЗКУ

**Проблематика.** Зображення з аномальною анатомією голови є проблемними для автоматичних методів сегментації, як результат, погане визначення на таких зображеннях зон уваги впливає на кількісні показники та візуальну оцінку перфузійних даних.

**Мета дослідження.** Ця стаття пропонує новий підхід для повністю автоматизованого та відносно точного визначення зон уваги на зображеннях перфузійної динамічно-сприйнятливої контрастної магнітно-резонансної томографії, який може бути застосований у перфузійному аналізі.

**Методика реалізації.** У запропонованому підході результат визначення перфузійних зон уваги являє собою бінарну маску, в якій нульові значення задають на зображенні положення повітря, екстрацеребральних тканин і спинномозкової рідини. Процес створення бінарної маски починається з проведення порогової фільтрації для пікселів низької інтенсивності, яким будуть відповідати нульові значення маски. Оптимальне значення порога знаходиться через отримання інформації щодо інтенсивності пікселів у зоні приблизного анатомічного розміщення мозку. Алгоритми заповнення отворів і вирощування бінарної ділянки використовуються для обробки помилок проведеної порогової фільтрації та створення сегмента, що відповідає тканинам мозку. Надалі визначається положення спинномозкової рідини проведенням порогової фільтрації пікселів високої інтенсивності з-поміж пікселів сегмента, що відповідає тканинам мозку. Кожне зображення часової серії перфузійних даних використовується для уточнення розміщення нульових значень маски, що задають на зображенні положення спинномозкової рідини.

**Результати дослідження.** Для оцінки запропонованого підходу результати визначення перфузійних зон уваги були порівняні з результатами мануальної сегментації, що була виконана досвідченими радіологами і взята за еталонний стандарт. Результати визначення перфузійних зон уваги на 120 зображеннях добре узгоджуються з еталонним стандартом з індексом Дайса 0,9576 ± 0,013 (чутливість і специфічність становлять 0,9931 ± 0,0053 і 0,9730 ± 0,0111 відповідно). Результати визначення всіх перфузійних зон уваги були визнані двома досвідченими радіологами як достатні для клінічного застосування.

**Висновки.** Результати показують, що запропонований підхід може бути застосований для виявлення зон уваги на зображеннях перфузійної динамічно-сприйнятливої контрастної магнітно-резонансної томографії. Сегментація на основі запропонованого підходу може бути реалізована як частина будь-якої автоматизованої системи обробки зображень мозку клінічного застосування.

**Ключові слова:** перфузійна динамічно-сприйнятлива контрастна магнітно-резонансна томографія; зрізи з аномальною анатомією мозку; зона уваги; сегментація; порогова фільтрація.

С.Н. Алхимова

АВТОМАТИЧЕСКОЕ ОПРЕДЕЛЕНИЕ ЗОН ИНТЕРЕСА НА МАГНИТНО-РЕЗОНАНСНЫХ ПЕРФУЗИОННЫХ ИЗОБРАЖЕНИЯХ МОЗГА

**Проблематика.** Изображения с аномальной анатомией головы являются проблемными для автоматических методов сегментации, как результат, плохое определение на таких изображениях зон интереса влияет на количественные показатели и визуальную оценку перфузионных данных.

**Цель исследования.** Эта статья предлагает новый подход для полностью автоматизированного и относительно точного определения зон интереса на изображениях перфузионной динамично-восприимчивой контрастной магнитно-резонансной томографии, который может быть применен в перфузионном анализе.

**Методика реализации.** В предложенном подходе результат определения перфузионных зон интереса представляет собой бинарную маску, в которой ее нулевые значения задают на изображении положения воздуха, экстрацеребральных тканей и спинномозговой жидкости. Процесс создания бинарной маски начинается с выполнения пороговой фильтрации для пикселей низкой интенсивности, которым впоследствии будут соответствовать нулевые значения маски. Оптимальное значение порога находится путем получения информации об интенсивности пикселей в зоне приблизительного анатомического расположения мозга. Алгоритмы заполнения отверстий и выращивания бинарной области используются для обработки ошибок проведенной пороговой фильтрации и создания сегмента, который соответствует тканям мозга. В дальнейшем определяется положение на изображении спинномозговой жидкости путем проведения пороговой фильтрации пикселей высокой интенсивности в сегменте, который соответствует тканям мозга. Каждое изображение часовой серии перфузионных данных используется для уточнения расположения нулевых значений маски, которые задают положение спинномозговой жидкости.

**Результаты исследования.** Для оценивания предложенного подхода было выполнено сравнение результатов определения перфузионных зон интереса с мануальной сегментацией, результаты которой были получены опытными радиологами и были приняты как эталонные. Определение перфузионных зон интереса предложенным подходом на 120 изображениях хорошо согласуется с эталонными данными с индексом Дайса 0,9576 ± 0,013 (чувствительность и специфичность составляют 0,9931 ± 0,0053 и 0,9730 ± 0,0111 соответственно). Результаты определения всех перфузионных зон интереса были признаны двумя опытными радиологами как достаточные для клинического применения.

**Выводы.** Результаты исследования показывают, что предложенный подход может быть применен для определения зон интереса на изображениях перфузионной динамично-восприимчивой контрастной магнитно-резонансной томографии. Сегментация на основе предложенного подхода может быть реализована как часть любой автоматизированной системы клинического применения для обработки изображений мозга.

**Ключевые слова:** перфузионная динамично-восприимчивая контрастная магнитно-резонансная томография; срезы с аномальной анатомией мозга; зона интереса; сегментация; пороговая фильтрация.